\documentclass[reprint, superscriptaddress, secnumarabic, amssymb, nobibnotes, aps, prl]{revtex4-1}

\usepackage{graphicx}
\usepackage{epstopdf}
\usepackage[T1]{fontenc}
\usepackage{amsbsy}
\usepackage{gensymb}
\usepackage[T1]{fontenc}
\usepackage{amsmath}
\usepackage{amssymb}
\usepackage{bbm}
\usepackage{braket}
\usepackage{xcolor}
\allowdisplaybreaks
\usepackage{graphicx}
\usepackage[colorlinks=true]{hyperref}  
\hypersetup{
    bookmarks=true,         % show bookmarks bar?
    unicode=false,          % non-Latin characters 
    pdftoolbar=true,        % show Acrobat
    pdfmenubar=true,        % show Acrobat 
    pdffitwindow=false,     % window fit to page when opened
    pdfstartview={FitH},    % fits the width of the page to the window
    pdftitle={Superconducting ground state of nonsymmorphic superconducting compound Zr$_{2}$Ir},    % title
    pdfauthor={Manasi Mandal},      %author
    pdfsubject={},   % subject of the document
    pdfcreator={},   % creator of the document
    pdfproducer={}, % producer of the document
    pdfkeywords={} {} {}, % list of keywords
    pdfnewwindow=true,      % links in new window
    colorlinks=true,       % false: boxed links; true: colored links
    linkcolor=blue, %red,          % color of internal links (change box color with linkbordercolor)
    citecolor=blue,        % color of links to bibliography
    filecolor=magenta,      % color of file links
    urlcolor=blue           % color of external links
} 

\usepackage[normalem]{ulem}

\newcommand{\figref}[1]{Fig.~\ref{#1}}

\newcommand{\tableref}[1]{Table~\ref{#1}}

\begin{document}
\title{\textrm{Superconducting ground state of nonsymmorphic superconducting compound Zr$_{2}$Ir}}
\affiliation{Department of Physics, Indian Institute of Science Education and Research Bhopal, Bhopal, 462066, India}
\author{Manasi~Mandal}
\affiliation{Department of Physics, Indian Institute of Science Education and Research Bhopal, Bhopal, 462066, India}
\author{Chandan Patra}
\affiliation{Department of Physics, Indian Institute of Science Education and Research Bhopal, Bhopal, 462066, India}
\author{Anshu Kataria}
\affiliation{Department of Physics, Indian Institute of Science Education and Research Bhopal, Bhopal, 462066, India}
\author{D.~Singh}
\affiliation{ISIS Facility, STFC Rutherford Appleton Laboratory, Harwell Science and Innovation Campus, Oxfordshire, OX11 0QX, UK}
\author{P.~K.~Biswas}
\affiliation{ISIS Facility, STFC Rutherford Appleton Laboratory, Harwell Science and Innovation Campus, Oxfordshire, OX11 0QX, UK}
\author{J.~S.~Lord}
\affiliation{ISIS Facility, STFC Rutherford Appleton Laboratory, Harwell Science and Innovation Campus, Oxfordshire, OX11 0QX, UK}
\author{A.~D.~Hillier}
\affiliation{ISIS Facility, STFC Rutherford Appleton Laboratory, Harwell Science and Innovation Campus, Oxfordshire, OX11 0QX, UK}
\author{R.~P.~Singh}
\email[]{rpsingh@iiserb.ac.in}
\affiliation{Department of Physics, Indian Institute of Science Education and Research Bhopal, Bhopal, 462066, India}

\date{\today}

\begin{abstract}

The nonsymmorphic Zr$_{2}$Ir alloy is a possible topological semimetal candidate material and as such may be part of an exotic class of superconductors. Zr$_{2}$Ir is a superconductor with a transition temperature of 7.4 K with critical fields of 19.6(3) mT and 3.79(3) T, as determined by heat capacity and magnetisation. Zero field muon spin relaxation measurements show that time-reversal symmetry is preserved in these materials. The specific heat and transverse field muon spin rotation measurements rule out any possibility to have a nodal or anisotropic superconducting gap, revealing a conventional s-wave nature in the superconducting ground state. Therefore, this system is found to be conventional nonsymmorphic superconductor, with time-reversal symmetry being preserved and an isotropic superconducting gap.

\end{abstract}
\maketitle

\section{Introduction}

Topological superconductors (TSCs) have emerged as an exciting exotic class of unconventional superconductors. These materials have a nontrivial topology with a superconducting gap in the bulk but with topologically protected states at the surfaces \cite{TSC1,TSC2,TI2,TI3,TI5,TI4}. These surface states can host Majorana fermions, which may lead to possible applications for in fault-tolerant quantum computation \cite{TSC1,TSC2}. The search for new TSCs is currently a central challenge in quantum materials research. However, despite tremendous research activity, only a handful of compounds are reported as a potential candidate to bulk topological superconductors, including Sr$_{2}$RuO$_{4}$ \cite{SrRuO,SrRuO2}, Au$_{2}$Pb \cite{AuPb}, PbTaSe$_{2}$ \cite{PbTaSe}, BiPd \cite{BiPd}, $\beta$-PdBi$_{2}$ \cite{PdBi}, and MoTe$_{2}$ \cite{MoTe}.\\

Recently, experimental studies performed on theoretically predicted topological binary and ternary superconducting compounds, such as: NbC, TaC \cite{NbC}, and A15 Ti$_{3}$X (X = Ir, Sb) \cite{A15_1,A15_2}, ZrRuAs \cite{ZrRuAs}, Kagome flat band LaRu$_{3}$Si$_{2}$ \cite{LaRu3Si2} have shown conventional s-wave superconductivity with preserved time-reversal symmetry (TRS). In contrast, multi-gap superconductivity has been reported in the topological superconducting candidate TaOsSi \cite{TaOsSi} and TRS is preserved. For RRuB$_{2}$ (R = Y, Lu), s-wave superconductivity with spin fluctuations has been reported\cite{RRuB2_1, RRuB2_2}. These studies failed to provide the role of the nontrivial surface state on superconducting ground states. The limited number of topological superconductors makes it challenging to determine the exact superconducting pairing mechanism and role of topological surface states on superconducting properties. It is essential to discover new materials or study the existing ones to look for superconductivity with topologically protected surface states.\\

In this paper, a detailed study of the superconducting properties of the nonsymmorphic alloy (space group $I4/mcm$, No. 140) Zr$_{2}$Ir using magnetic susceptibility, electrical resistivity, heat capacity and muon-spin rotation/relaxation ($\mu$SR) techniques. Recent work has revealed that the nonsymmorphic symmetry can host novel topological phases \cite{NSym} and topological superconductivity, thus making Zr$_2$Ir an ideal candidate material. Initial band structure calculations have revealed that Zr$_{2}$Ir is a topological semi-metal with a symmetry enforced Fermi level degeneracy at high symmetry points \cite{topo1,topo2}. A symmetry-enforced topological semi-metal may hold low-energy excitations and reveal novel topological response phenomena and unusual magneto-transport properties \cite{topo_semi}. Several bands close to the Fermi surface with distinct electron masses can lead to different electron-phonon coupling strengths, which results in different gap energies. This can promote an exotic superconducting ground state depending on the strength of inter and intra-band coupling \cite{multigap}. So, it is very intriguing and timely to study the superconducting ground state and time-reversal symmetry in Zr$_{2}$Ir alloy. Our results show that Zr$_{2}$Ir can be described as fully gapped s-wave order parameters and preserved time-reversal symmetry in the superconducting state. Interestingly, the Uemura plot suggests Zr$_{2}$Ir lie in the vicinity of Zr$_{3}$Ir and other unconventional superconductors, which break time-reversal symmetry \cite{NbOs2,Z3I,Z3I2}.
    
\section{Experimental Details}

A polycrystalline sample of Zr$_{2}$Ir was made using a stoichiometric mixture of Zr (99.99 $\% $) and Ir (99.97 $\% $) by standard arc melting method under a high purity argon gas atmosphere. The sample was flipped and remelted several times to improve the chemical homogeneity. The resulting 'button' had negligible mass loss during this process, and was annealed at 750 $\degree$C for three days in vacuum-sealed quartz tube. Room temperature powder X-ray powder diffraction (XRD) data was collected using a PANalytical X$^{,}$pert Pro diffractometer using Cu $K_{\alpha}$ radiation ($\lambda$ = 1.54056 $\text{\AA}$). The DC susceptibility measurements were performed using a MPMS3 (Quantum Design). Resistivity and specific heat measurements were performed in the Physical Property Measurement System (PPMS, Quantum Design). The $\mu$SR measurements were carried out at the ISIS pulsed neutron and muon source at the STFC Rutherford Appleton Laboratory, United Kingdom using the MuSR spectrometer. Zero-ﬁeld muon spin relaxation (ZF-$\mu$SR) measurements were performed in the temperature range 0.3 - 10 K. The transverse-ﬁeld muon spin rotation (TF-$\mu$SR) measurements were performed at 60 mT, well above the lower critical field H$_{C1}$(0). We have analyzed the $\mu$SR data \cite{DOI} using the Mantid software.

\begin{figure}[htbp!]
\includegraphics[width=1.0\columnwidth]{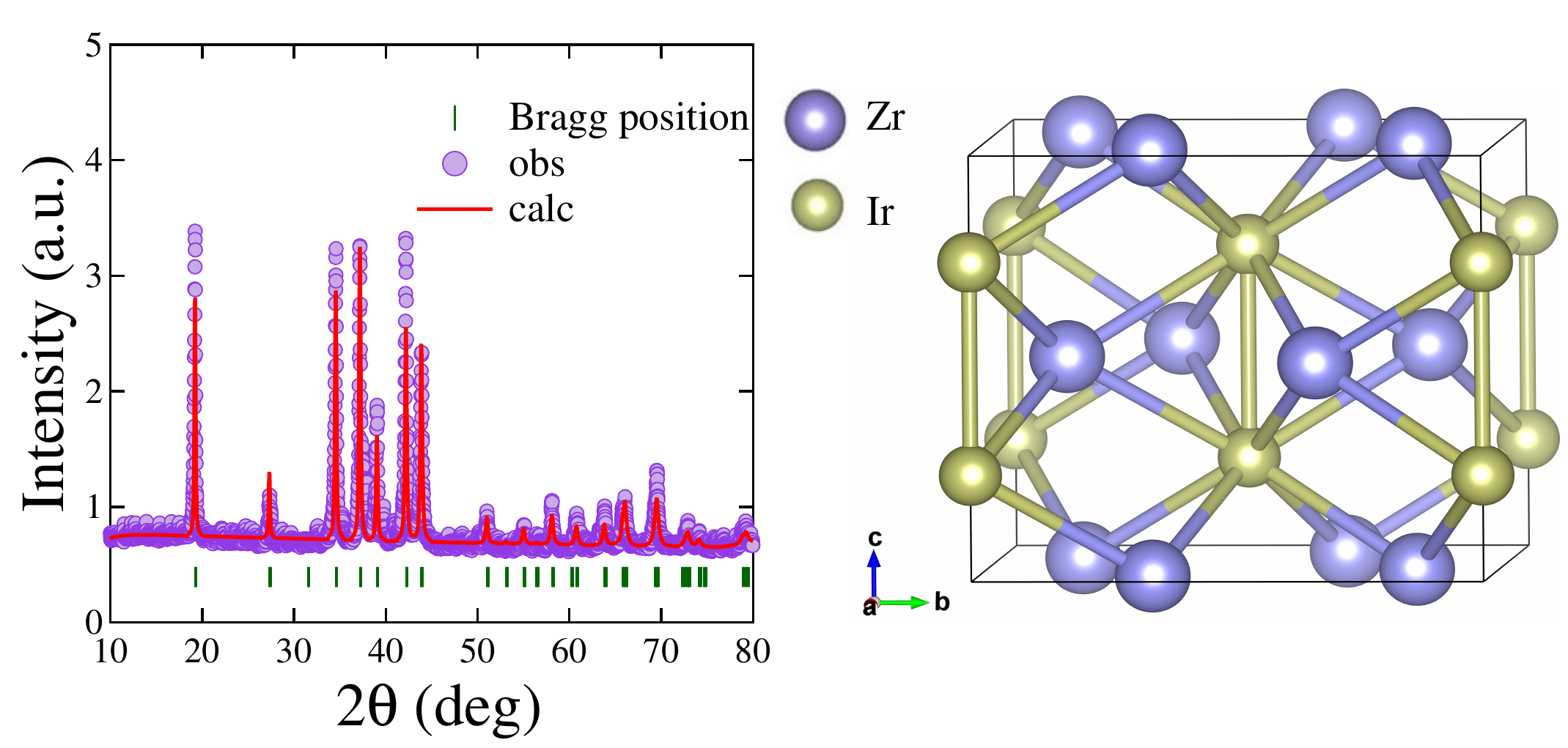}
\caption {\label{Fig1} (a) Room temperature XRD pattern of Zr$_{2}$Ir shows the phase purity. (b) CuAl$_{2}$ type tetragonal structure of Zr$_{2}$Ir with space group $I4/mcm$.}
\end{figure}

\section{Results and Discussions}

Rietveld refinement of the room temperature XRD pattern was performed using the FullProf software package \cite{Full_Prof}. This confirmed the phase purity of the sample with negligible Zr, or any other, impurity. The crystal structure was confirmed as the tetragonal CuAl$_{2}$ type with space group $I4/mcm$ (space group no - 140). The refined XRD pattern is shown in \figref{Fig1}. The refined lattice parameters are summarized in \tableref{tbl:lattice_parameters}.

\begin{table}[htbp!]
\caption{Crystal structure parameters of Zr$_{2}$Ir}
Space group $I4/mcm$ (no - 140)\\
$\alpha$ = $\beta$= $\gamma$ = 90$\degree$
\label{tbl:lattice_parameters}
\setlength{\tabcolsep}{12pt}
\begin{center}
\begin{tabular}[b]{|l| c| c| }\hline 
 Parameters & Unit & Zr$_{2}$Ir\\
\hline\hline
%\\[0.1ex]                                
a &$\text{\AA}$ & 6.51(6)\\
c &$\text{\AA}$ & 5.66(1)\\
V$_{cell}$& $\text{\AA}^{3}$&  240.70 (4) \\
%\\[0.1ex]
\hline
\end{tabular}
\par\medskip\footnotesize
\end{center}

\begin{center}
\setlength{\tabcolsep}{8pt}
\begin{tabular}[b]{l c c c c}\hline
Atom & Wyckoff position & x & y & z\\
\hline\hline
Zr & 2a & 0.0 & 0.0 & 0.0 \\
Ir & 6c & 0.25 & 0.0 & 0.50\\
\hline
\end{tabular}
\par\medskip\footnotesize
\end{center}
\end{table}

\begin{figure}[htbp!]
\includegraphics[width=1.0\columnwidth]{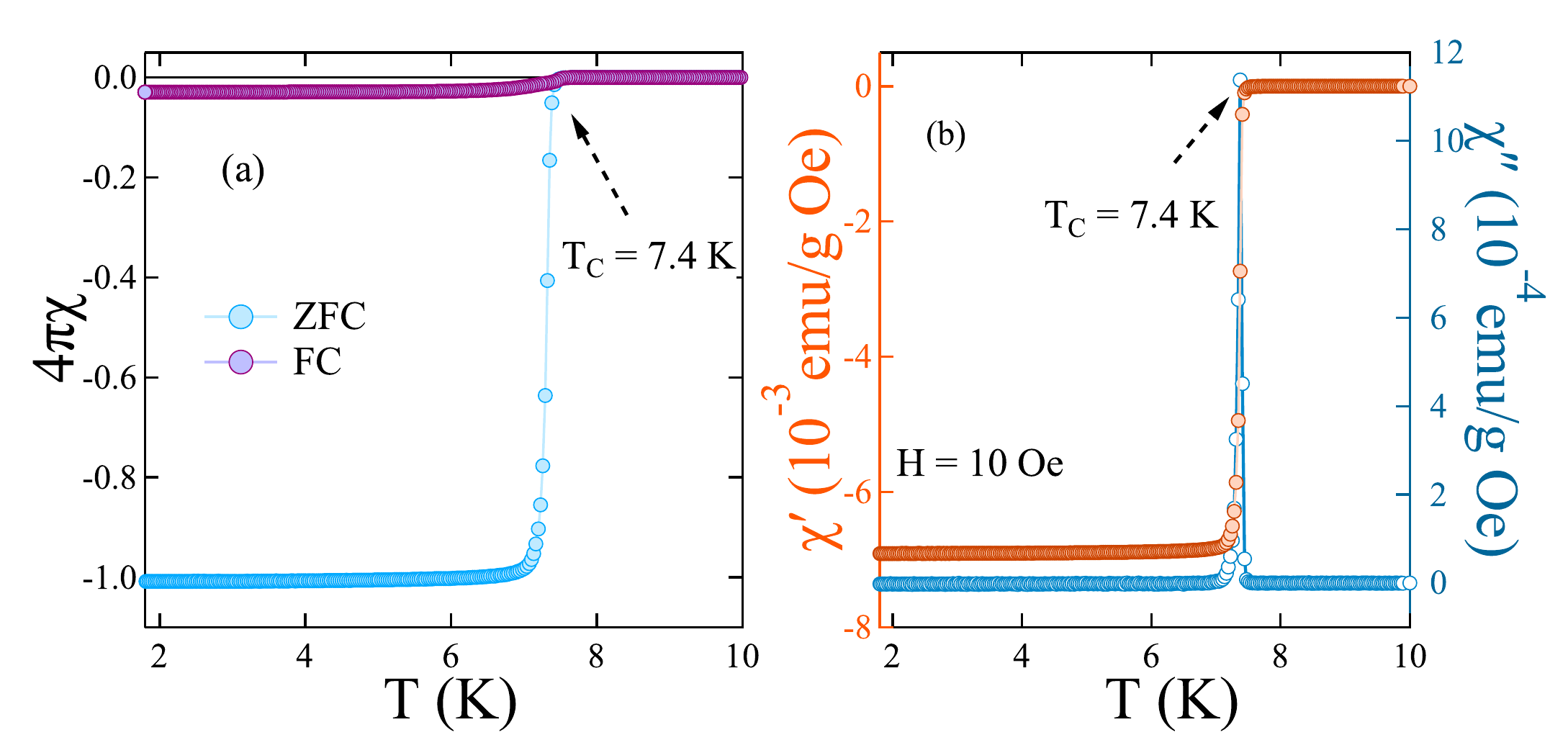}
\caption {\label{Fig2} (a) The temperature dependence of the dc magnetization in an applied field of 1 mT and shows bulk superconductivity with sharp diamagnetic transition at T$_{C}$ = 7.4 K. (b) The temperature dependence of the ac susceptibility and further confirms superconductivity at 7.4 K.}
\end{figure}

The magnetization was measured in both zero-field cooled (ZFC) and field cooled (FC) mode with an applied field, H = 1 mT. These measurement confirms bulk superconductivity at T$_{C}$ = 7.4(1) K (\figref{Fig2}(a)) with 100$\%$ superconducting volume fraction. AC susceptibility measurements further confirm the superconductivity at 7.4 K and is shown in \figref{Fig2}(b).

\begin{figure}[htbp!]
\includegraphics[width=1.0\columnwidth]{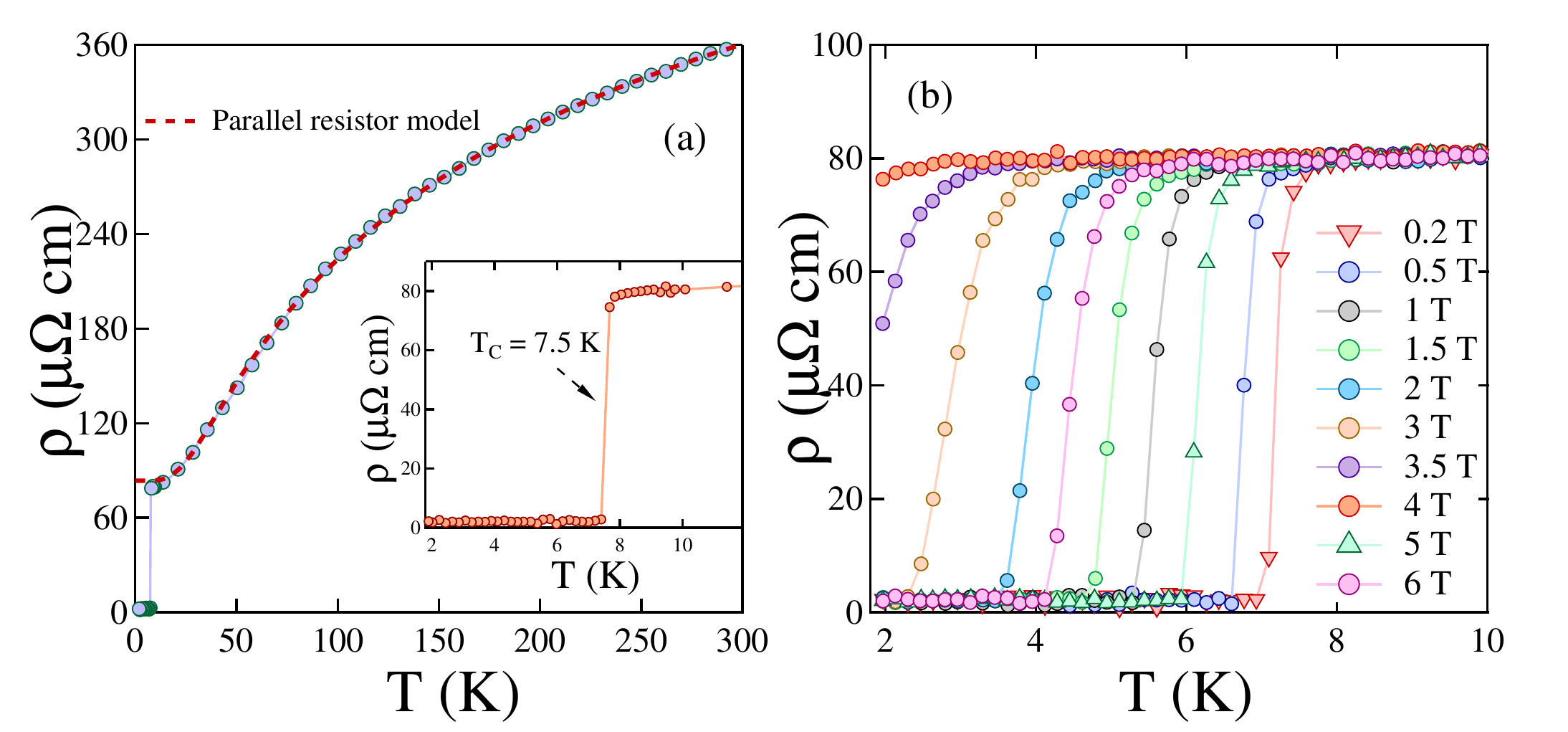}
\caption {\label{Fig3} (a) Temperature dependence of the resistivity and the inset shows a drop to zero resistivity. (b) Temperature dependence of the resistivity with different applied magnetic fields.}
\end{figure}

The temperature dependence of the resistivity were measured in a range of different applied fields, from zero to 6 T, and is shown in \figref{Fig3}. It also confirms the presence of superconductivity (inset of \figref{Fig3} (a)). The normal state resistivity data (10 K $\le$ T $\le$ 300 K) is fitted using the parallel resistor model \cite{parallel,Zr} given by 
\begin{equation}
\frac{1}{\rho(T)}=\frac{1}{\rho_{1}(T)}+\frac{1}{\rho_{sat}}
\label{eq1:RES1}
\end{equation}
where $\rho_{sat}$ is the high-temperature saturation resistivity, and $\rho_{1}$(T) is the ideal temperature-dependent resistivity given by the following expression

\begin{equation}
\rho_{1}(T)=\rho_{0}+r\left(\frac{T}{\theta_R}\right)^{5}\int_{0}^{\theta_R/T} \frac{x^5}{(e^{x}-1)(1-e^{-x})}dx
\label{eqn2:RES2}
\end{equation}

where the second term is due to inelastic electron-phonon scattering. $\theta_{D}$ is the Debye temperature, r is a pre-factor and depends on the electronic structure of metal through the Fermi velocity and density of state, and $\rho$(0) is the residual resistivity. The best fit yields a residual resistivity $\rho$(0) = 98(1) $\mu\ohm$ cm, a Debye temperature $\theta_{D}$ = 148(3) K, and a high temperature saturation resistivity $\rho_{sat}$ = 557(5) $\mu\ohm$ cm. The value of $\theta_{D}$ is close to that obtained from the specific heat data (described later).

\begin{figure}[htbp!]
\includegraphics[width=1.0\columnwidth]{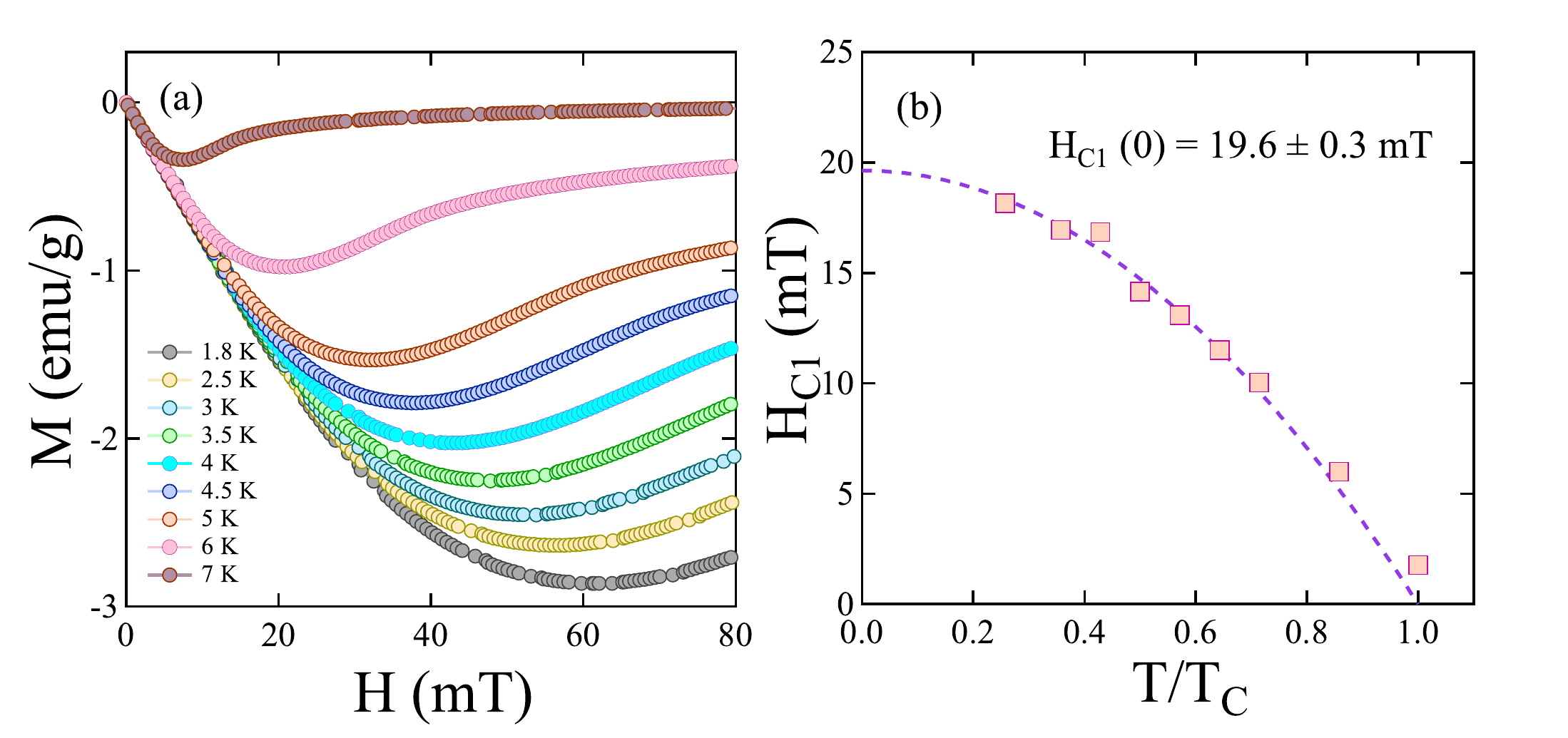}
\caption {\label{Fig4} (a) Low field magnetization curves at different temperatures. (b) The temperature dependence of the lower critical field is shown and the line is a fit to the data using the GL equation \eqref{eqn1:HC2}.} .
\end{figure}

To find the lower critical field, H$_{C1}$(0), the low field magnetization were collected at different temperatures, as shown in \figref{Fig4} (a). We have fitted the temperature dependence of H$_{C1}$ with GL-equation $H_{C1}(T)=H_{C1}(0)\left(1-\left(\frac{T}{T_{C}}\right)^{2}\right)$and estimated H$_{C1}$(0) as 19.6(3) mT (see \figref{Fig4} (b)).

\begin{figure}[htbp!]
\includegraphics[width=1.0\columnwidth]{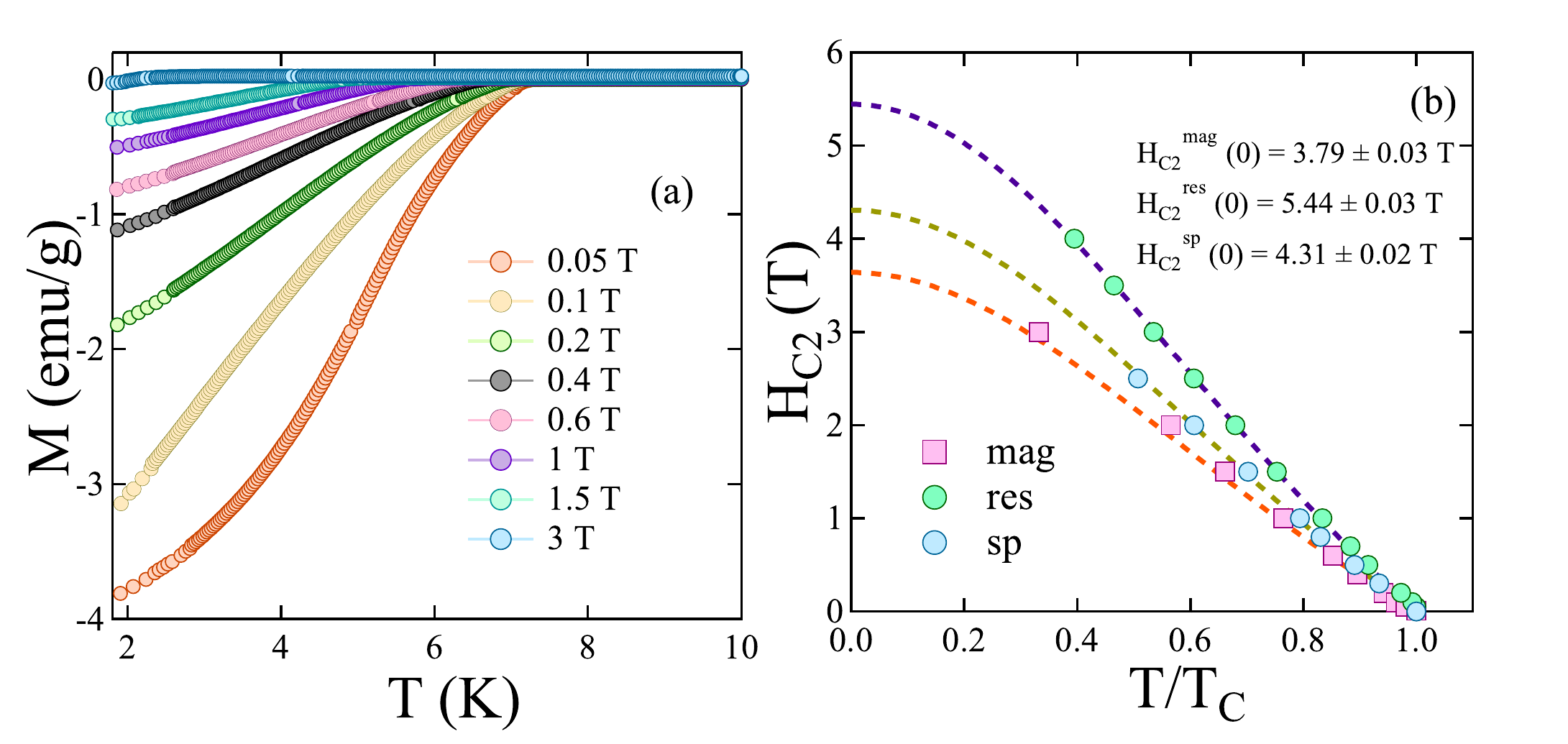}
\caption {\label{Fig5} (a) The magnetization data were collected at different fields in the temperature range of 1.8 K $\le$ T $\le$ 10 K. (b) The temperature dependence of the upper critical field, H$_{C2}$(T) was determined from the magnetization, specific heat, and resistivity data collected at different fields, with Ginzburg-Landau (GL) formula.}
\end{figure}

The magnetization was also measured with different magnetic fields over a temperature range of 1.8 K $\le$ T $\le$ 10 K, in order to determine the upper critical field, H$_{C2}$(0). Values for H$_{C2}$(T) were determined by taking the onset temperature as the criteria for T$_{C}$. \figref{Fig5} (b) shows the upper critical field as a function of temperature. The data were fitted using the Ginzburg-Landau (GL) formula

\begin{equation}
H_{C2}(T) = H_{C2}(0)\frac{(1-t^{2})}{(1+t^2)}
\label{eqn1:HC2}
\end{equation}
where t = T/T$_{C}$ and gives H$_{C2}^{mag}$(0) = 3.79(3) T. We have also calculated H$_{C2}$(T) by using the same relation \eqref{eqn1} from the specific heat and resistivity data collected at different fields, which gives slightly higher H$_{C2}$(0) value \figref{Fig5}(b). 

We have calculated the coherence length, $\xi_{GL}(0)$ as 93.2(1) \text{\AA} using the relation $H_{C2}(0) = \frac{\Phi_{0}}{2\pi\xi_{GL}^{2}}$ and H$_{C2}^{mag}$(0) = 3.79 T, where $\Phi_{0}$ (= 2.07 $\times$10$^{-15}$ T m$^{2}$) is the magnetic flux quantum \cite{tin}. Another fundamental length scale of a superconductor is the magnetic penetration length $\lambda_{GL}$(0) can be determined using the equation \eqref{eqn8:lamda}. The calculated value of $\lambda_{GL}(0)$ is 1700(20).

\begin{equation}
H_{C1}(0) = \frac{\Phi_{0}}{4\pi\lambda_{GL}^2(0)}\left( ln \frac{\lambda_{GL}(0)}{\xi_{GL}(0)}+0.497\right)  
\label{eqn8:lamda}
\end{equation}

The GL-parameter, $k_{GL}$ = $\frac{\lambda_{GL}(0)}{\xi_{GL}(0)}$ = 18.2(2) and shows the alloy is a type-II superconductor.\\

\begin{figure}[htbp!]
\includegraphics[width=1.0\columnwidth]{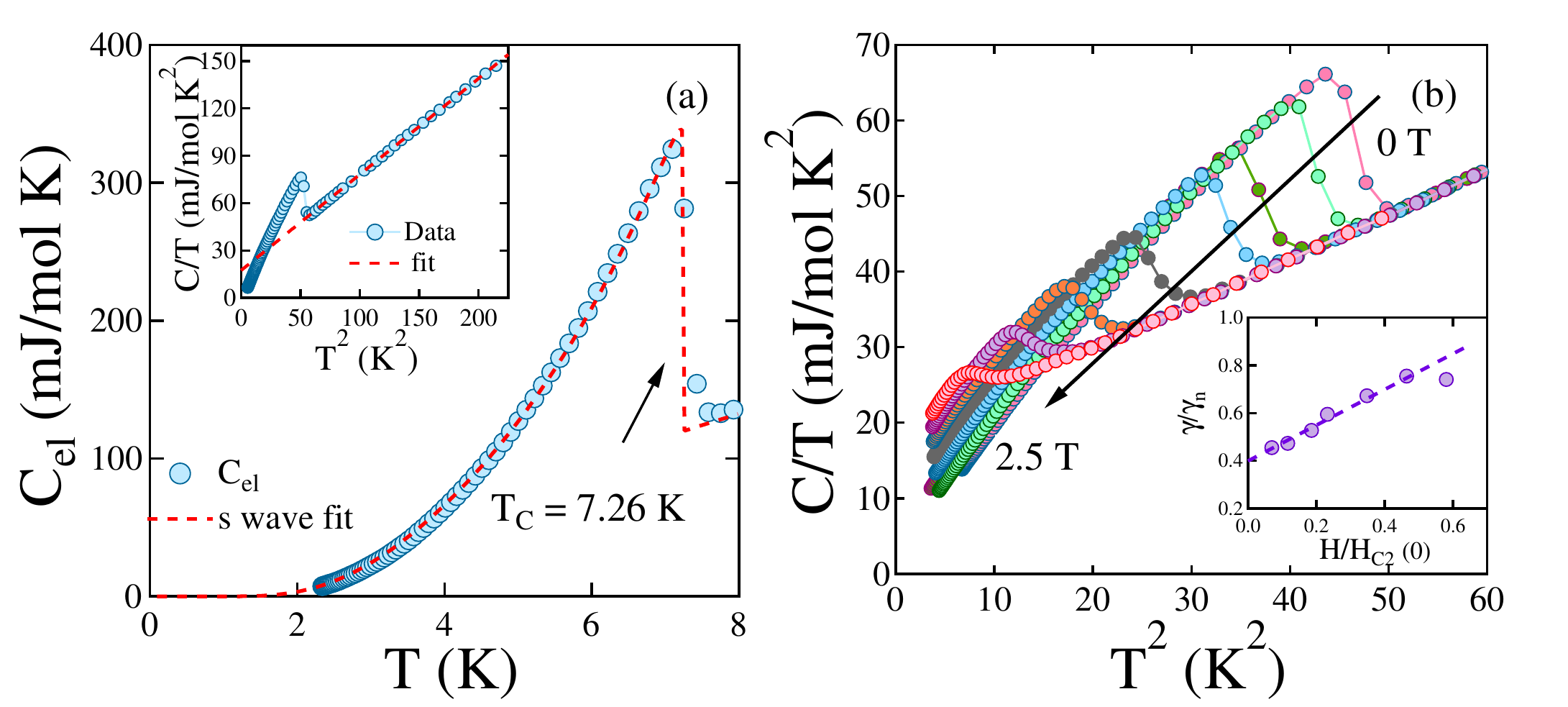}
\caption {\label{Fig6} (a) The temperature dependence of C$_{el}$. The line is the result of the fit to an isotropic s-wave model. $\frac{C}{T}$ vs T$^{2}$ is shown in inset, where the line is the result of the fit from the relation $\frac{C}{T}$ = $\gamma_{n}$ + $\beta$T$^2$. (b) The temperature dependence C/T at different fields is shown. The inset shows the field variation of $\gamma$ and is clearly shown to be linear.}
\end{figure}

To find the electron-phonon co-relation and the nature of superconductivity in Zr$_{2}$Ir, the specific heat was measured in zero field and in different applied magnetic fields. A sharp jump at T = 7.26(1) K verifies bulk superconductivity. \figref{Fig6} (b) shows $\frac{C}{T}$ vs T$^{2}$ data, where the normal state data was fitted using the relation $\frac{C}{T}$ = $\gamma_{n}$ + $\beta$T$^2$. $\gamma_{n}$ and $\beta$ are Sommerfeld coefficient and Debye constant respectively, which signifies electron-electron and electron-phonon co-relation. The fit yields $\gamma_{n}$ = 17.4(2) mJ/mol K$^2$ and $\beta_{}$= 0.61(2) mJ/mol K$^4$. Electronic contribution C$_{el}$ was extracted from the total specific heat using the relation C$_{el}$ = C - $\beta$T$^3$. The C$_{el}$ data is best described by a conventional s-wave model and gives a superconducting gap value of $\frac{\Delta(0)}{k_{B}T_{C}}$ = 1.99(1) indicating an isotropic superconducting gap at the Fermi surface \cite{BCS}. The obtained superconducting gap value is greater than the BCS predicted value (1.76). We have extracted C$_{el}$ for different applied magnetic fields and the data was fitted for each field using the relation $\frac{C_{el}}{T}$ = $\gamma + \frac{A}{T}exp\left(\frac{-bT_{C}}{T}\right)$ \cite{sas} and is shown in \figref{Fig6}(b). The field variation of the Sommerfeld coefficient $\gamma$ is linear, i.e. $\gamma$ $\propto$ H, and further indicates an isotropic gap \cite{line_nodes,line_nodes2}.

The Debye temperature $\theta_{D}$ was calculated using the relation $\theta_{D}$ = $\left(\frac{12\pi^{4}RN}{5\beta_{3}}\right)^{\frac{1}{3}}$ where N (= 3) is the number of atoms per formula unit, R is the molar gas constant (= 8.314 J mol$^{-1}$ K$^{-1}$) and was found to be 209(2) K.
The density of states at the Fermi level $D_{C}(E_{F})$ is 8.0(1) $\frac{states}{eV f.u}$ and was determined from the relation $\gamma_{n}$ = $\left(\frac{\pi^{2}k_{B}^{2}}{3}\right)D_{C}(E_{f})$, where k$_{B}$ is Boltzman constant. The strength of the attractive interaction between the electrons and phonons ($\lambda_{e-ph}$) was calculated using the McMillan equation \cite{McMillan} and was found to be 0.83(1) indicating that Zr$_{2}$Ir is a moderately coupled superconductor similar to superconducting topological candidate materials, such as: Ti$_{3}$Ir and Ti$_{3}$Sb \cite{A15_2}. All the superconducting and normal state parameters are summarized in \tableref{tbl:parameters}.

\begin{figure}[htbp!]
\includegraphics[width=1.0\columnwidth]{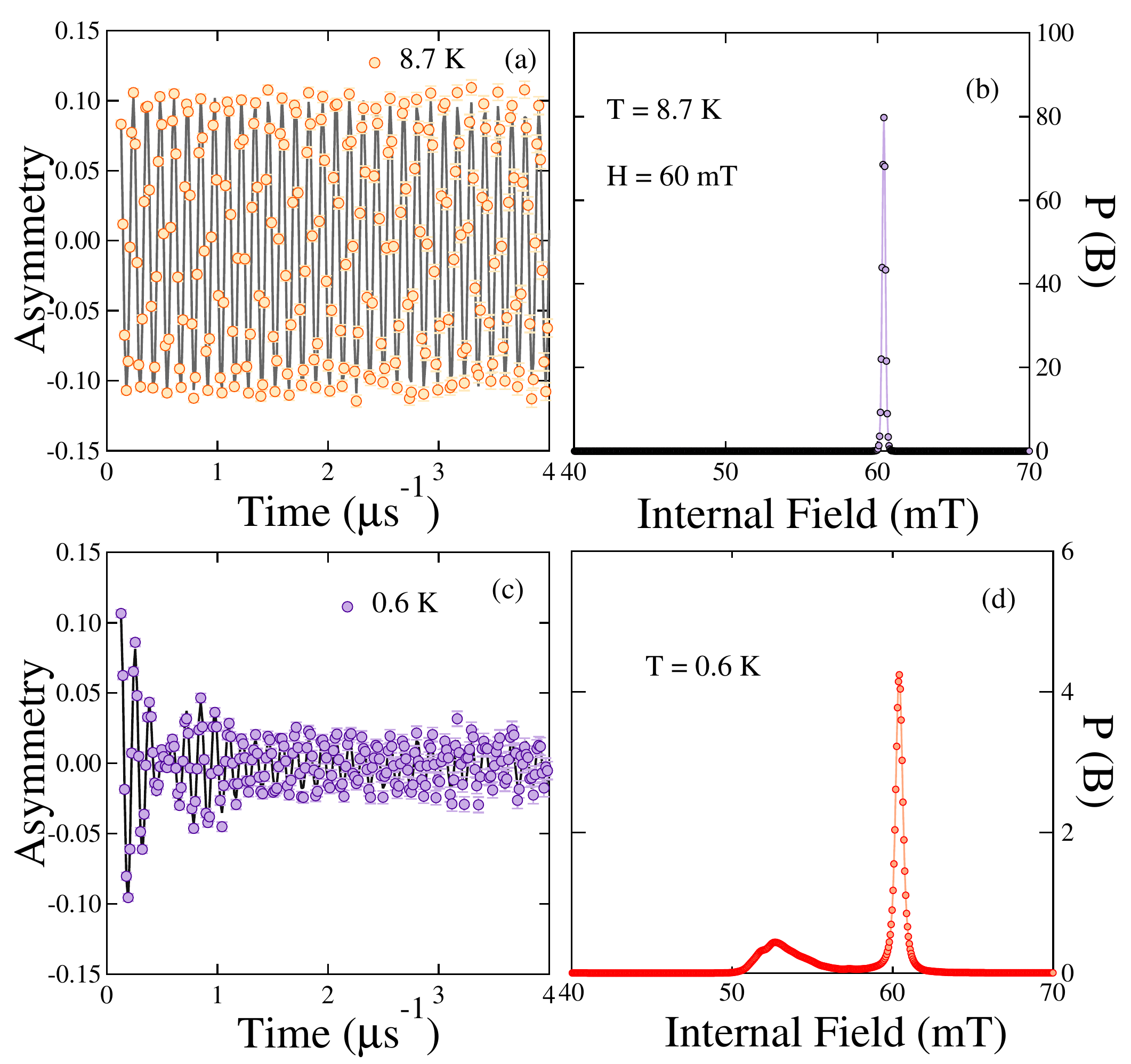}
\caption{\label{Fig7} (a) and (c) TF-$\mu$SR spectra are shown above and below superconducting transition temperature at 60 mT, whereas (b) and (d) show the corresponding field distribution.}
\end{figure}

To further explore the superconducting gap structure, transverse field muon spin rotation (TF-$\mu$SR) spectroscopy measurements were taken at different temperatures with an applied field of 60 mT. The applied field ensures a well-ordered flux line lattice at temperatures below T$_{C}$ as the applied field is greater than H$_C$(0). \figref{Fig7} (a) and (c) shows the TF-$\mu$SR spectra above and below the superconducting transition temperature. The typical Maximum Entropy analysis results of the magnetic field distribution, extracted from the TF-$\mu$SR time spectra, are shown in the \figref{Fig7} (b) mixed state, and (d) normal state \cite{Maxent}. The normal state spectra shows a homogeneous field distribution, whereas the spectra in the superconducting state show strong depolarization. This nature specifies the formation of inhomogeneous field distribution in the flux line lattice (FLL) state.

\begin{figure}[htbp!]
\includegraphics[width=1.0\columnwidth]{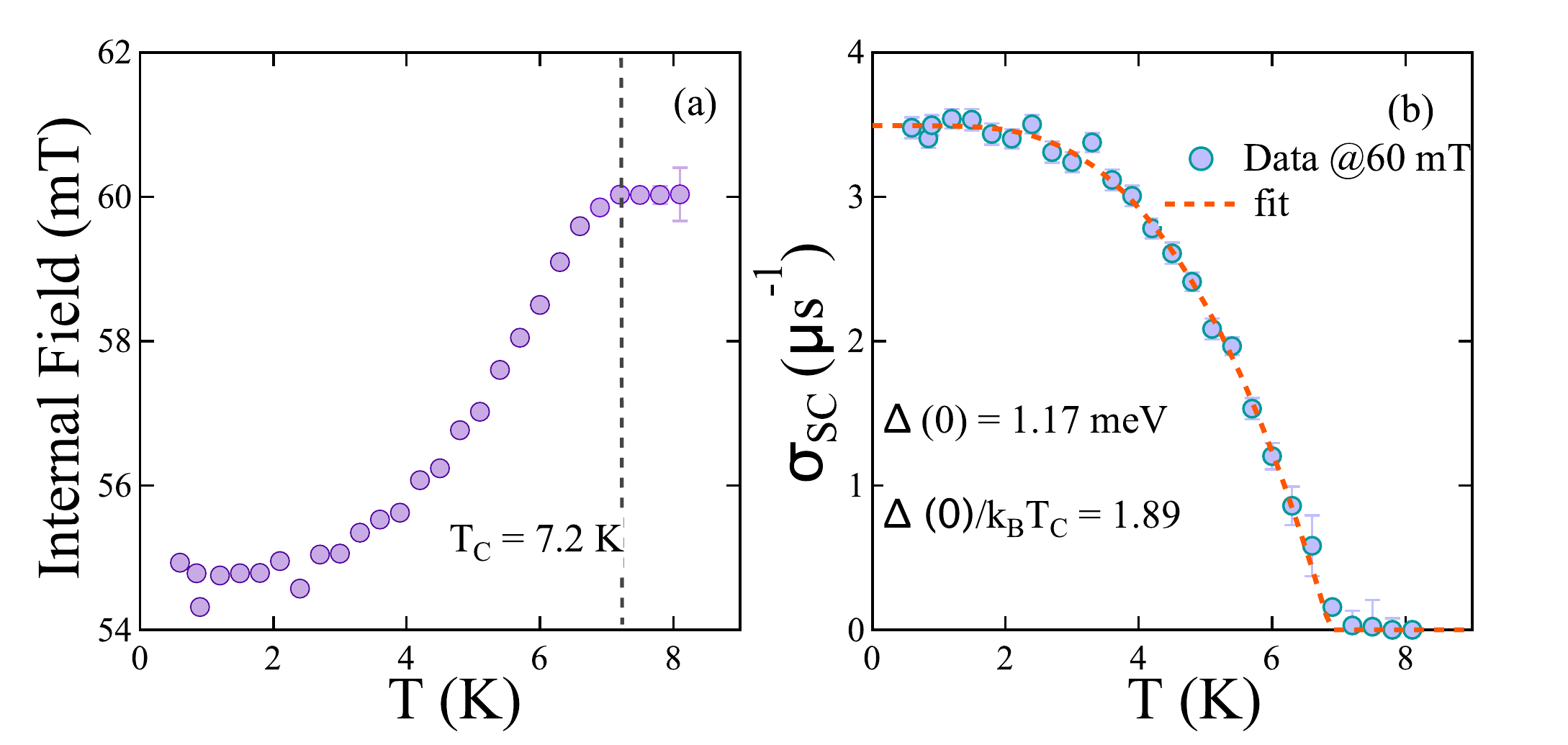}
\caption {\label{Fig8} (a) The temperature dependence of the internal magnetic field experienced by the muon ensemble (b) The temperature dependence of the muon depolarization rate with an applied field of 60 mT, whereas the dotted red line is the result of the fit to the data using a clean s-wave model.}
\end{figure}

The decaying Gaussian oscillatory function (G$_{TF}$(t)) best fit the TF-spectra, which consists of two decay components \eqref{eqn2}. An oscillatory background term was also added to account for the effect of the muons implanted directly into the silver sample holder which does not depolarize \eqref{eqn2}. A$_{bg}$ and B$_{bg}$ are the background contributions for the asymmetry and the field, respectively.

\begin{eqnarray}
G_{\mathrm{TF}}(t) &=&
\sum_{\mathrm{i}=1}^{2} A_{\mathrm{i}}\mathrm{exp}\left(\frac{-\sigma_{\mathrm{i}}^{2}t^{2}}{2}\right)\mathrm{cos}(\gamma_{\mu}B_{\mathrm{i}}t+\phi)\nonumber\\&+&A_{bg}\mathrm{cos}(\gamma_{\mu}B_{bg}t+\phi)
\label{eqn2}
\end{eqnarray}

where $\phi$ is the offset of the muon spin polarization with respect to positron detector and $\gamma_{\mu}$/2$\pi$ = 135.5 MHz/T is muon gyromagnetic ratio. The temperature dependence of effective depolarization rate $\sigma$ is related to the first and second moments by the relations \eqref{eqn3} and \eqref{eqn4_1} \cite{second_moment}

\begin{eqnarray}
\mathrm{<}B\mathrm{>} &=& \sum_{\mathrm{i}=1}^{2}\frac{A_{\mathrm{i}}\mathrm{B_{i}}}{A_{1}+A_{2}}
\label{eqn3}
\end{eqnarray}

\begin{eqnarray}
\mathrm{<}\Delta B^{2}\mathrm{>} = \sum_{\mathrm{i}=1}^{\mathrm{2}}\frac{A_{\mathrm{i}}[(\sigma_{\mathrm{i}}/\gamma_{\mu})^2+(B_{\mathrm{i}}-\mathrm{<}B\mathrm{>})^2]}{A_{\mathrm{1}}+A_{\mathrm{2}}}= \frac{\sigma^2}{\gamma_{\mu}^2}
\label{eqn4_1}
\end{eqnarray}

\figref{Fig8} (a) shows the temperature dependence of internal magnetic field $\mathrm{<}B\mathrm{>}$. The flux expulsion at T$_{C}$ is clearly evident from the reduction of the internal magnetic field $\mathrm{<}B\mathrm{>}$. The temperature variation of effective depolarization rate, $\sigma$ was extracted by fitting the TF-spectra which is consist of contribution from nuclear dipolar moments ($\sigma_{\mathrm{N}}$) and field variation across the flux line lattice ($\sigma_{\mathrm{sc}}$). We have extracted $\sigma_{\mathrm{sc}}$ by the quadratic relation $\sigma^{2}$ = $\sigma_{\mathrm{sc}}^{2}+\sigma_{\mathrm{N}}^{2}$. The $\sigma_{\mathrm{sc}}$ increases systematically with decreasing temperature and saturates at the lowest temperatures and is shown in \figref{Fig8} (b). This result suggests that Zr$_3$Ir does not exhibit line or point node and has an isotropic gap, in agreement with the heat capacity results. The data were fitted within the local London approximation for a BCS superconductor in the clean limit \cite{LaPtSi} which yields the superconducting energy gap $\Delta(0)$ is 1.17(1) meV and $ \frac{\Delta(0)}{k_{B} T_{C}} $ is 1.88(2), greater than the BCS predicted value 1.76. This value is consistent with the specific heat measurement.

\begin{figure}[htbp!]
\includegraphics[width=1.0\columnwidth]{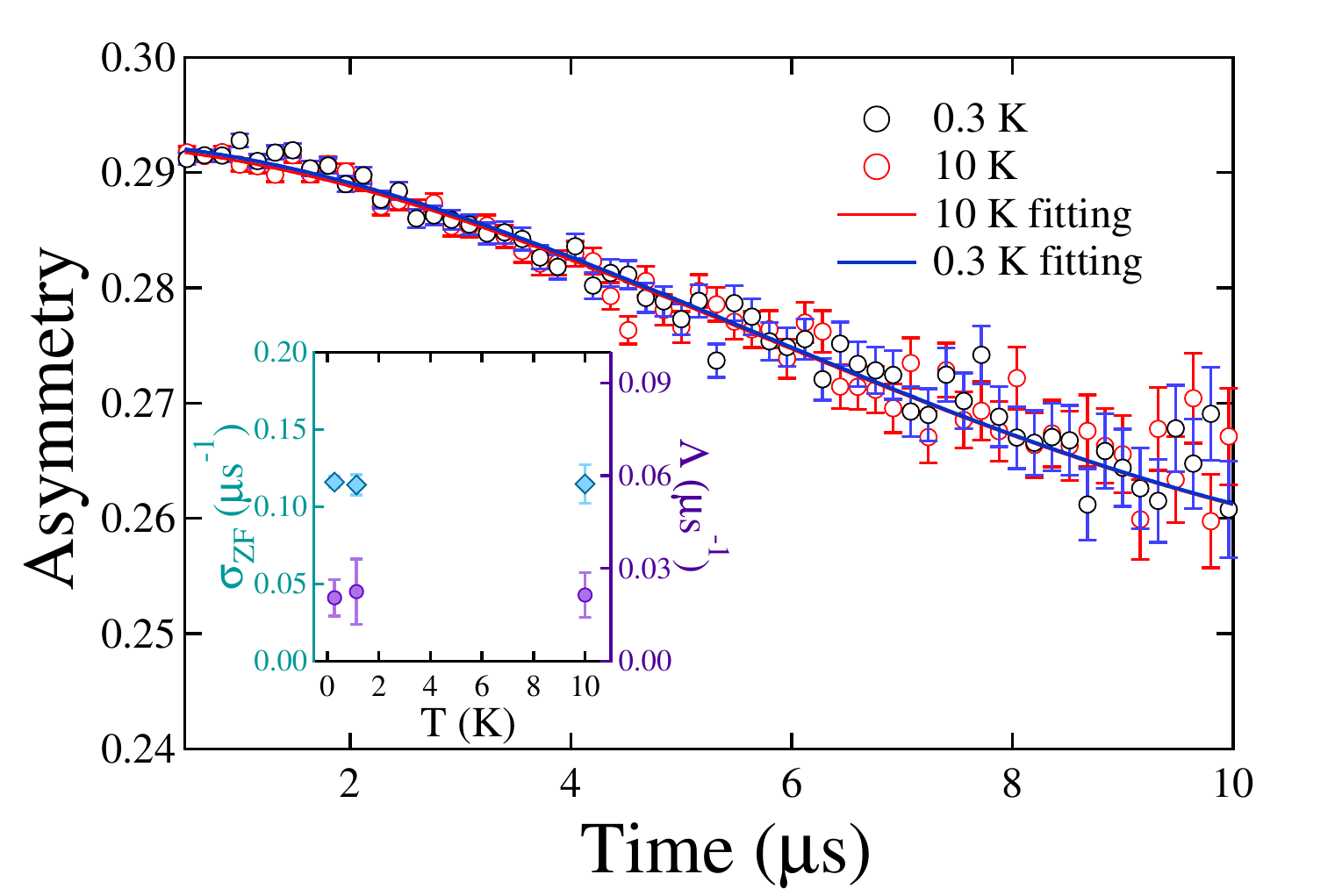}
\caption {\label{Fig9} Zero-field $\mu$SR spectra collected below (0.3 K) and above (10 K) the superconducting transition temperature. The solid blue line is the fit to static Kubo-Toyabe times exponential decay function.}
\end{figure}

In a type-II isotropic superconductor with a hexagonal Abrikosov vortex lattice having $\kappa$ $\mathrm{>}$ 5 and H/H$_{C2}$ $\leq$ 0.25 with penetration depth $\lambda$ to be calculated to a high degree of accuracy by the relation \cite{lambda1,lambda2}
\begin{equation}
\sigma_{\mathrm{sc}} [\mu s^{-1}] = 4.854\times 10^{4}(1-h)[1+1.21(1-\sqrt{h})^{3}]\lambda^{-2} [nm^{-2}]
\label{eqn4}
\end{equation}
where h = $ H/H_{C2}(T)$ is the reduced field. $\lambda_{GL}(0)$ was found to be 1169(10) \text{\AA} using H$_{C2}$(0) = 3.79 T.\\

To investigate the presence of a spontaneous magnetic field in the sample, we have collected ZF-$\mu$SR spectra at temperatures above and below the superconducting transition temperature. The asymmetry spectra were best fitted by static Kubo-Toyabe function multiplied by an exponential decay component \cite{kubo} is given by
\begin{equation}
G(t)= A_{1}\mathrm{exp}(-\Lambda t)G_{\mathrm{KT}}(t)+A_{\mathrm{BG}} ,
\label{eqn6}
\end{equation}
with \begin{eqnarray}
G_{\mathrm{KT}}(t) &=&\frac{1}{3}+\frac{2}{3}(1-\sigma^{2}_{\mathrm{ZF}}t^{2})\mathrm{exp}\left(-\frac{\sigma^{2}_{\mathrm{ZF}}t^{2}}{2}\right),
\label{eqn7}
\end{eqnarray}
where $A_{1}$ is the initial sample asymmetry, $\sigma_{\mathrm{ZF}}$ and $\Lambda$ are the Gaussian and an additional relaxation rate, respectively. $A_{\mathrm{BG}}$ is background contribution associated with muon stopping in the silver sample holder.
\figref{Fig9} shows no change in observed zero-field asymmetry spectra within the detection limit of $\mu$SR. This confirms the absence of any spontaneous magnetic field and a preserved time reversal symmetry in Zr$_{2}$Ir. 

\begin{figure}[htbp!]
\includegraphics[width=1.0\columnwidth]{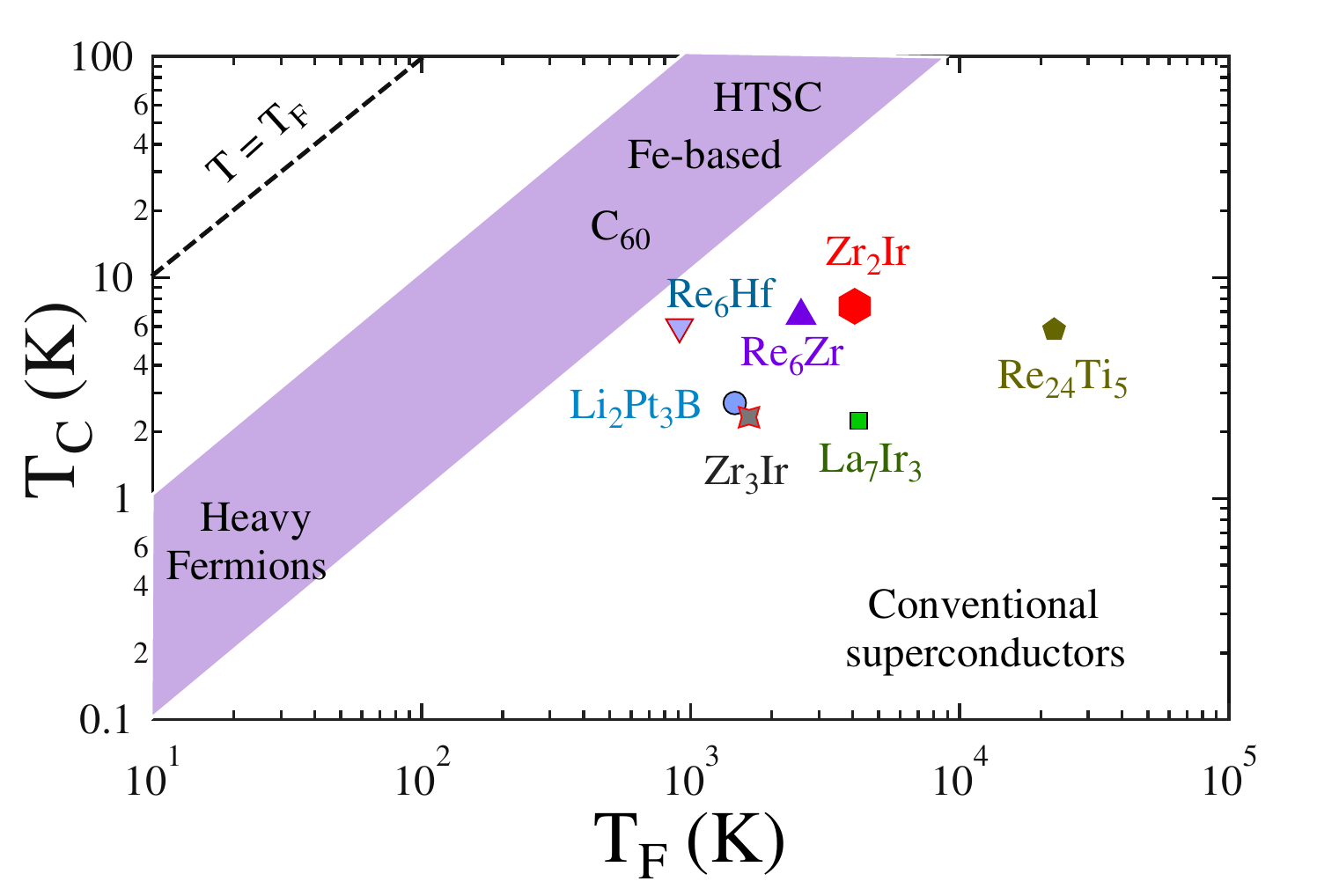}
\caption {\label{Fig10} The Uemura plot showing the superconducting transition temperature T$_{C}$ with respect to the effective Fermi temperature T$_{F}$, where Zr$_{2}$Ir is shown as a solid red marker. The violet solid band are the different families of unconventional superconductors plotted with some other unconventional superconductors \cite{NbOs2,Z3I}.}
\end{figure}

Uemura et al. \cite{Uemura1,Uemura2,Uemura3,Uemura4} have described a method of classifying unconventional superconductors based on the ratio of $\frac{T_{C}}{T_F}$, where T$_{F}$ is Fermi temperature. We have extracted T$_{F}$ by solving a set of five equations simultaneously as done in \cite{A15_2,NbOs2}. The estimated value of T$_{F}$ is 4069 K that gives $\frac{T_{C}}{T_F}$ as 0.002. This ratio falls out of the unconventional band 0.01$\leq$ $\frac{T_{C}}{T_F}$ $\leq$ 0.1 as predicted by the Uemura classification scheme but close to some other unconventional superconductors \cite{NbOs2,Z3I}, as shown in \figref{Fig10}.

\begin{table}[htbp!]
\caption{Normal and superconducting parameters of Zr$_{2}$Ir }
\setlength{\tabcolsep}{12pt}
\begin{center}
\label{tbl:parameters}
\begin{tabular}[b]{|l|c|c|c|}\hline
Parameters& Unit& Zr$_{2}$Ir   \\
\hline
T$_{C}^{mag}$& K & 7.4(1)   \\

H$_{C1}(0)$ & mT & 19.6(3)  \\ 

H$_{C2}^{mag}$(0) & T & 3.79(3)   \\

H$_{C2}^{res}$(0) & T & 5.44(2)   \\

$\xi_{GL}$ & \text{\AA} & 93.2(1)   \\

$\lambda_{GL}^{mag}$ & \text{\AA}& 1700(20)  \\

$\lambda_{GL}^{muon}$(0) & \text{\AA}& 1169(10) \\

$k_{GL}$ & & 18.2(2)   \\

$\gamma_{n}$ & mJ/mol K$^{2}$ & 17.4(2)  \\

$\beta$ & mJ/mol K$^{4}$  & 0.61(1) \\

$\frac{\Delta C}{\gamma T_{C}}$ &   &  1.53(2) \\

$\frac{\Delta(0)}{k_{B}T_{C}}$ &   &  1.99(1) (specific heat)  \\

$\frac{\Delta(0)}{k_{B}T_{C}}$ &   &  1.88(2) (muon) \\

$\theta_{D}$ & K & 209(2) (specific heat)  \\

$\theta_{D}$ & K & 148(3) (resistivity)  \\

$\lambda_{e-ph}$ &  & 0.83(1)  \\

D$_{C}$(E$_{f}$) & states/eV f.u. &  8.0(1) \\

T$_{F}$ & K &  4069(25)\\

\hline
\end{tabular}
\par\medskip\footnotesize
\end{center}
\end{table} 

\section{Conclusion}

We report superconductivity in possible topological semimetal Zr$_{2}$Ir having nonsymmorphic crystal structure. It shows the bulk type-II superconductivity with transition temperature T$_{C}$ = 7.4(1) K with the lower and upper critical field as 19.6(3) mT and 3.79(3) T, respectively. Specific heat measurements exhibit that superconducting energy gap, $\frac{\Delta_{0}}{k_{B}T_{C}}$ is 1.99(1), greater than BCS predicted value (1.76). Transverse field $\mu$SR measurements further confirm the superconducting energy gap as s-wave. Zero fields $\mu$SR measurements reveal the preserved time reversal symmetry in the superconducting ground state. This compound can be concluded as TRS preserved nonsymmorphic material. This work encourages further investigations on nonsymmorphic materials to understand the role of crystal symmetry and TRS on the possible bulk topological superconductivity.

%\vspace*{0.5cm}

\section{Acknowledgments} R.~P.~S.\ acknowledges the Science and Engineering Research Board, Government of India for the Core Research Grant CRG/2019/001028 and Financial support from DST-FIST Project No. SR/FST/PSI-195/2014(C) is also thankfully acknowledged. We thank ISIS, STFC, United Kingdom, for the muon beam time \cite{DOI}.


\begin{thebibliography}{100}

\bibitem{TSC1} X. L. Qi and S. C. Zhang, Rev. Mod. Phys. 83, 1057 (2011).

\bibitem{TSC2} J. Wang, Natl. Sci. Rev. 6, 2 (2019).

\bibitem{TI2} X. L. Qi and S. C. Zhang, Rev. Mod. Phys. 83, 1057 (2011).

\bibitem{TI3} C. Nayak, S. H. Simon, A. Stern, M. Freedman, and S. D.Sarma, Rev. Mod. Phys. 80, 1083 (2008).

\bibitem{TI5} L. Fu and C. L. Kane, Phys. Rev. Lett. 100, 096407 (2008).

\bibitem{TI4} N. Read and D. Green, Phys. Rev. B 61, 10267 (2000).

\bibitem{SrRuO} Y. Maeno, S. Kittaka, T. Nomura, S. Yonezawa, and K. Ishida, J. Phys. Soc. Jpn. 81, 011009 (2012).

\bibitem {SrRuO2} G. M. Luke,Y. Fudamoto, K. M.Kojima, M. I. Larkin, J.Merrin, B. Nachumi, Y. J. Uemura, Y. Maeno, Z. Q. Mao, Y. Mori, H. Nakamura, and M. Sigrist, Nature (London) 394, 558 (1998).

\bibitem{AuPb} Y. Xing, H. Wang, C.-K. Li, X. Zhang, J. Liu, Y. Zhang, J. Luo, Z. Wang, Y. Wang, L. Ling, M. Tian, S. Jia, J. Feng, X.-J. Liu, J. Wei, and J. Wang, Nature 16005, 5 (2016).

\bibitem{PbTaSe} G. Bian, T.-R. Chang, R. Sankar, S.-Y. Xu, H. Zheng, T. Neupert, C.-K. Chiu, S.-M. Huang, G. Chang, I. Belopolski, D. S. Sanchez, M. Neupane, N. Alidoust, C. Liu, B. Wang, C.-C. Lee, H.-T. Jeng, C. Zhang, Z. Yuan, S. Jia, A. Bansi, F. Chou, H. Lin, and M. Z. Hasan, Nat. Commun. 10556, 7 (2016).

\bibitem{BiPd} Z. Sun, M. Enayat, A. Maldonado, C. Lithgow, E. Yelland, D. C. Peets, A. Yaresko, A. P. Schnyder, and P. Wahl, Nat. Commun. 6633, 6 (2015).

\bibitem{PdBi} P. K. Biswas, D. G. Mazzone, R. Sibille, E. Pomjakushina, K. Conder, H. Luetkens, C. Baines, J. L. Gavilano, M. Kenzelmann, A. Amato, and E. Morenzoni, Phys. Rev. B 93, 220504(R) (2016).

\bibitem{MoTe} Y. Qi, P. G. Naumov, M. N. Ali, C. R. Rajamathi, O. Barkalov, Y. Sun, S. Chandra, S.-C. Wu, V. Su$\beta$, M. Schmidt, P. Echhard, P. Werner, R. Hillebrand, F. Tobias, E. Kampertt, W. Schnelle, S. Parkin, R. J. Cava, C. Felser, B. Yan, S. A. Medvedev, Nat. Commun. 7, 11038 (2016).

\bibitem{NbC} T. Shang, J. Z. Zhao, D. J. Gawryluk, M. Shi, M. Medarde, E. Pomjakushina, and T. Shiroka, Phys. Rev. B 101, 214518 (2020).

\bibitem{A15_1} M. Kim, C. Wang, and K. Ho, Phys. Rev. B 99, 224506 (2019).
	
\bibitem{A15_2} M. Mandal, Sajilesh K. P., R. Roy Chowdhury, D. Singh, P. K. Biswas, A. D. Hillier, and R. P. Singh, Phys. Rev. B 103, 054501 (2021).

\bibitem{ZrRuAs} D. Das, D. T. Adroja, M. R. Lees, R. W. Taylor, Z. S. Bishnoi, V. K. Anand, A. Bhattacharyya, Z. Guguchia, C. Baines, H. Luetkens, G. B. G. Stenning, L. Duan, X. Wang, and C. Jin, Phys. Rev. B 103, 144516 (2021).

\bibitem{LaRu3Si2} C. Mielke, Y. Qin, J.-X. Yin, H. Nakamura, D. Das, K. Guo, R. Khasanov, J. Chang, Z. Q. Wang, S. Jia, S. Nakatsuji, A. Amato, H. Luetkens, G. Xu, M. Z. Hasan, and Z. Guguchia, Phys. Rev. Materials 5, 034803 (2021).

\bibitem{TaOsSi} C. Q. Xu, B. Li, J. J. Feng, W. H. Jiao, Y. K. Li, S. W. Liu, Y. X. Zhou, R. Sankar, N. D. Zhigadlo, H. B. Wang, Z. D. Han, B. Qian, W. Ye, W. Zhou, T. Shiroka, P. K. Biswas, X. Xu, and Z. X. Shi, Phys. Rev. B 100, 134503 (2019).

\bibitem{RRuB2_1} J. A. T. Barker, R. P. Singh, A. D. Hillier, and D. McK. Paul, Phys. Rev. B 97, 094506 (2018).

\bibitem{RRuB2_2} Y. Gao, P.-J. Guo, K. Liu, and Z.-Y. Lu, Phys. Rev. 102, 115137 (2020).

\bibitem{NSym} Y.-H. Chan, B. Kilic, M. M. Hirschmann, C.-K. Chiu, L. M. Schoop, D. G. Joshi, and A. P. Schnyder, Phys. Rev. Mater. 3, 124204 (2019).

\bibitem{topo1} B. Bradlyn, L. Elcoro, J. Cano, M. G. Vergniory, Z. Wang, C. Felser, M. I. Aroyo, and B. A. Bernevig, Nature 547, 298 (2017).

\bibitem{topo2} M. G. Vergniory, L. Elcoro, C. Felser, N. Regnault, B. A. Bernevig, and Z. Wang, Nature 566, 480 (2019).

\bibitem{topo_semi} Y. X. Zhao and A. P. Schnyder, Phys. Rev. B 94, 195109 (2016).

\bibitem{multigap} H. Suhl, B. T. Matthias, and L. R. Walker, Phys. Rev. Lett. 3, 552 (1959).

\bibitem{NbOs2} D. Singh, Sajilesh K. P., S. Marik, A. D. Hillier, and R. P. Singh, Phys. Rev. B 99, 014516 (2019).

\bibitem{Z3I} Sajilesh K. P., D. Singh, P. K. Biswas, Gavin B. G. Stenning, A. D. Hillier, and R. P. Singh, Phys. Rev. Mater. 3, 104802 (2019).

\bibitem{Z3I2} T. Shang, S. K. Ghosh, J. Z. Zhao, L.-J. Chang, C. Baines, M. K. Lee, D. J. Gawryluk, M. Shi, M. Medarde, J. Quintanilla, and T. Shiroka, Phys. Rev. B 102, 020503(R) (2020).

\bibitem{DOI} 10.5286/ISIS.E.RB2010460.

\bibitem{Full_Prof} J. Rodriguez-Carvajal, Physica B. 192, 55 (1993).

\bibitem{Zr} D. A. Mayoh, J. A. T. Barker, R. P. Singh, G. Balakrishnan, D. McK. Paul, and M. R. Lees, Phys, Rev. B 96, 064521 (2017).

\bibitem{parallel} H. Wiesmann, M. Gurvitch, H. Lutz, A. K. Ghosh, B. Schwarz, M. Strongin, P. B. Allen, and J. W. Halley, Phys. Rev. Lett. 38, 782 (1977).

\bibitem{tin} M. Tinkham, $\textit{Introduction to Superconductivity}$, 2nd ed. (McGraw - Hill, New York, 1996). 

\bibitem{BCS} H. Padamsee, J. E. Neighbor, C. A. Shiffman, J. Low. Temp. Phys. 12, 387 (1973).

\bibitem{sas} M. Isobe, A. Masao, and N. Shirakawa, Phys. Rev. B 93, 054519 (2016).

\bibitem{line_nodes} N. Nakai, P. Miranovic, M. Ichioka, and K. Machida, Phys. Rev. B 70, 100503(R) (2004).

\bibitem{line_nodes2} M. Isobe, M. Arai, and N. Shirakawa, Phys. Rev. B 93, 054519 (2016).

\bibitem{McMillan} W. L. McMillan, Phys. Rev. 167, 331 (1968).

\bibitem{Maxent} B. D. Rainford and G. J. Daniell, Hyperfine Interact. 87, 1129 (1994).

\bibitem{second_moment} M. Weber, A. Amato, F. N. Gygax, A. Schenck, H. Maletta, V. N. Duginov, V. G. Grebinnik, A. B. Lazarev, V. G. Olshevsky, V. Yu. Pomjakushin, S. N. Shilov, V. A. Zhukov, B. F. Kirillov, A. V. Pirogov, A. N. Ponomarev, V. G. Storchak, S. Kapusta, and J. Bock, Phys. Rev. B 48, 13022 (1993).

\bibitem{LaPtSi} Sajilesh K. P., D. Singh, A. D. Hillier, and R. P. Singh, Phys. Rev. B 102, 094515 (2020).

\bibitem{lambda1} E. H. Brandt, J. Low Temp. Phys. 73, 355 (1988).

\bibitem{lambda2} E. H. Brandt, Phys. Rev. B 68, 054506 (2003).

\bibitem{kubo} R. S. Hayano, Y. J. Uemura, J. Imazato, N. Nishida, T. Yamazaki, and R. Kubo, Phys. Rev. B 20, 850 (1979).

\bibitem{Uemura1} Y. J. Uemura, V. J. Emery, A. R. Moodenbaugh, M. Suenaga, D. C. Johnston, A. J. Jacobson, J. T. Lewandowski, J. H. Brewer, R. F. Kiefl, S. R. Kreitzman, G. M. Luke, T. Riseman, C. E. Stronach, W. J. Kossler, J. R. Kempton, X. H.
Yu, D. Opie, and H. E. Schone, Phys. Rev. B 38, 909(R) (1988).

\bibitem{Uemura2} Y. J. Uemura, G. M. Luke, B. J. Sternlieb, J. H. Brewer, J. F. Carolan, W. N. Hardy, R. Kadono, J. R. Kempton, R. F. Kiefl, S. R. Kreitzman, P. Mulhern, T. M. Riseman, D. L. Williams, B. X. Yang, S. Uchida, H. Takagi, J. Gopalakrishnan, A. W. Sleight, M. A. Subramanian, C. L. Chien, M. Z. Cieplak, G. Xiao, V. Y. Lee, B. W. Statt, C. E. Stronach, W. J. Kossler, and X. H. Yu, Phys. Rev. Lett. 62, 2317 (1989).

\bibitem{Uemura3} Y. J. Uemura, L. P. Le, G. M. Luke, B. J. Sternlieb, W. D. Wu, J. H. Brewer, T. M. Riseman, C. L. Seaman, M. B. Maple, M. Ishikawa, D. G. Hinks, J. D. Jorgensen, G. Saito, and H. Yamochi, Phys. Rev. Lett. 66, 2665 (1991).

\bibitem{Uemura4} A. D. Hillier and R. Cywinski, Appl. Magn. Reson. 13, 95 (1997).


\end{thebibliography}
\end{document}